%% file: main.tex
\documentclass[sigconf]{acmart}
\usepackage{booktabs} 
\usepackage{graphicx}
\usepackage{subcaption}
\usepackage{caption}
\usepackage{xspace}
\usepackage{url}

\setcopyright{rightsretained}

\acmDOI{10.475/123_4}

\acmISBN{123-4567-24-567/08/06}

\acmConference[WWW'19]{The Web Conference}{May 2019}{San Francisco, CA, USA} 
\acmYear{2019}
\copyrightyear{2019}

\acmArticle{4}
\acmPrice{15.00}

\newcommand{\chenhao}[1]{\textbf{\color{blue}[** #1 **]}}

\newcommand{\dr}[1]{\textbf{\color{purple}/* #1 (dr) */}}

\newcommand{\daj}[1]{\textbf{\color{orange}/* #1 (daj) */}}

\newcommand{\toc}[1]{\textbf{\color{red}[*toc* #1 *toc*]}}

\newcommand{\para}[1]{\noindent{\bf #1}\xspace}
\newcommand{\secref}[1]{Section~\ref{#1}\xspace}
\newcommand{\figref}[1]{Figure~\ref{#1}\xspace}

\newcommand{\communityname}[1]{{\sf #1}\xspace}

\copyrightyear{2019}
\acmYear{2019} 
\setcopyright{iw3c2w3}
\acmConference[WWW '19]{Proceedings of the 2019 World Wide Web Conference}{May 13--17, 2019}{San Francisco, CA, USA}
\acmBooktitle{Proceedings of the 2019 World Wide Web Conference (WWW '19), May 13--17, 2019, San Francisco, CA, USA}
\acmPrice{}
\acmDOI{10.1145/3308558.3313689}
\acmISBN{978-1-4503-6674-8/19/05}

\ccsdesc[300]{Human-centered computing~User studies}

\keywords{Online Communities; Success; Group Dynamics; Reddit}

\begin{document}
\title{
Are All Successful Communities Alike?
Characterizing and Predicting the Success of Online Communities
}
\renewcommand{\shorttitle}{Characterizing and Predicting the Success of Online Communities}


%



\author{Tiago Cunha}
\affiliation{University of Michigan}
\email{tiolivei@umich.edu}

\author{David Jurgens}
\affiliation{University of Michigan}
\email{jurgens@umich.edu}

\author{Chenhao Tan}
\affiliation{University of Colorado Boulder}
\email{chenhao.tan@colorado.edu}

\author{Daniel M. Romero}
\affiliation{University of Michigan}
\email{drom@umich.edu}





\begin{abstract}
The proliferation of online communities has created exciting opportunities to study the mechanisms that explain group success.
While a growing body of research 
investigates community success through a single measure --- typically, the number of members --- we argue that there are multiple ways of measuring success.
Here, we present a systematic study to understand the relations between these success definitions and test how well they can be predicted based on community properties and behaviors from the earliest period of a community's lifetime. 
%
We identify four success measures that are desirable for most communities: (i) growth in the number of members; (ii) retention of members; (iii) long term survival of the community; and (iv) volume of activities within the community. Surprisingly, we find that our measures do not exhibit very high correlations, suggesting that they capture different types of success. Additionally, we find that different success measures are predicted by different attributes of online communities, suggesting that success can be achieved through different behaviors. Our work sheds light on the basic understanding on what success represents in online communities and what predicts it. Our results suggest that success is multi-faceted and cannot be measured nor predicted by a single measurement. This insight has practical implications for the creation of new online communities and the design of platforms that facilitate such communities. 


\end{abstract}

\maketitle

\input{intro.tex}
\input{data.tex}
\input{framework.tex}

\input{prediction-features.tex}

\input{experiment.tex}

\input{limitations.tex}

\input{recommendations.tex}

\input{related.tex}

\input{conclusion.tex}

\begin{acks}
This research was partly supported by the National Science Foundation under Grant No. IIS-1617820.
\end{acks}

\bibliographystyle{ACM-Reference-Format}
\bibliography{refs}

\end{document}

%% file: intro.tex
\section{Introduction}

Understanding the complex ways in which communities are successful over time is essential for building and maintaining vibrant online communities \citep{kraut:2012,Kim:2000:CBW:518514}. 
User-created online communities in social media platforms enable people with shared points of view and interest to connect with each other and are one of the main ways in which the Web has 
shaped how people interact and find each other.
Although the low barrier for creating online communities has led to a large number of communities that are constantly involving and continually emerging,
not all of them are successful.
Fortunately, massive datasets from these online communities 
have allowed researchers to 
study the dynamics of the creation and lifecycles of online communities at a large scale \citep{Kairam:2012,Zhu:2014:IMO:2556288.2557213,Ducheneaut:ProceedingsOfChi:2007, tan:18, Backstrom:2006, centola:2010}. 
In particular, a problem of interest is to predict the eventual success of a new community from a set of community attributes that can be measured early in its lifetime. 
That is, {\em can we tell early on whether a community will be successful?}

An implicit yet important question is concerned with the definition of community success.
In fact, most existing research takes a very narrow view of what defines success in online communities. Existing research has typically considered a single measure of community success and this measure is often determined by the number of users who eventually participate in the community~\cite{Kairam:2012, tan:18, Backstrom:2006, centola:2010}. While this is a reasonable measure of success for many communities, it is not necessarily appropriate for all. Consider, for example, a community created with the goal of maintaining social relationships among the graduates from a small college on a given year. The members of such community would probably not define success by the size of their group, relative to other communities. A better measure of success in this case would be whether active members continue to stay active over time (i.e., retention). In contrast, a community created with the goal of raising awareness about an issue such as global warming would likely measure its success by how many people joined the cause. 

We argue that, given the rich and complex nature of 
online communities, considering a single measure of success has limited our understanding on what makes communities successful. Our work aims to fill this gap by exploring a variety of success measures of newly created communities on Reddit. 
We consider four classes of success measures that should be generally desirable by most communities: (i) growth in the number of members; (ii) retention of members; (iii) long term survival of the community; and (iv) volume of activities within the community. We find that, while these measures are positively correlated, the correlations are not very high, suggesting that each of the measures captures a fundamentally different aspect of a community's success. By focusing only on a community's growth in the number of members, we miss other dimensions that are also and sometimes even uniquely desirable. 

Given the observed differences between our success measures, we then explore the features of newly created communities that are predictive of future success. Since we expect that different measures of success be predicted by different features, we consider a variety of features to capture different aspects of the community early in its lifetime. In our framework, we observe a community from the time of its creation until the time when it has attracted a fixed number of members. Then, using attributes of the community that can be measured up until that time, we predict the success of the community based on each of our success measures. 

Specifically, we consider six different classes of features to predict success based on existing research and theory on online community success, group development, and organization theory:  (i) the volume and speed of initial activities; (ii) the distribution of activities over users and time;
(iii) the composition of its early members; (iv) linguistic style features of early content; (v) features of the communication network among early members; and (iv) the communities that early members belong to before they join the new community. We test the predictive performance of each class of features separately as well as all features combined. We find that no single class of features outperforms the combination of all regardless of the measure of success we use, suggesting that each class captures a meaningful dimension of the community's success. 
Furthermore, we find that the single most predictive class highly depends on the measure of success. For example, while linguistic style features are the best predictors of a community's survival, the distribution of activities over time best predicts retention. 

While we find that different measures of success are predicted by different classes of features, we also identify 
community behaviors that predict success across different measures. A particularly robust example is the inequality in the distribution of posts and comments per user. We find that successful communities tend to distribute their activities in a highly skewed fashion, where a small number of users 
contribute a large fraction of the activity in the community. Our interpretation of this pattern is that for a community to be successful, it needs to have a small number of committed members that maintain the community active early in its lifetime until the community reaches a critical mass. 
These observations resonate with the importance of 
levels of activity and commitment of an online community's founders for its ultimate success \citep{kraut:2014}.  

Understanding and predicting the success of online communities early in their lifetime can provide hypotheses for experimental studies to identify causal relations and help community creators manage their communities in ways that are known to predict success. Our findings suggest that online communities can be successful in a variety of ways and are predicted by different early behaviors. Thus, depending on the goals of a community, its members should focus on different aspects of their behavior in order to maximize their chances of success. 
Our results can help us begin to understand success in online communities as a multi-faceted concept and identify the behaviors that drive each of its possible facets.

%% file: data.tex
\section{Reddit Communities}

We begin with a brief description of 
our testbed, Reddit. Reddit is a popular social news website and forum, which allows users to self-organize into user-created communities by areas of interest called subreddits. In Reddit, most communities are public and users can join as many public communities as they desire. Reddit also allows users to submit content, such as textual posts or URLs to other contexts, comment on posts, and upvote and downvote content. We consider a user a member of a community if they post or comment in that community.

\begin{figure}[t]
    \centering
   \includegraphics[width=0.45\textwidth]{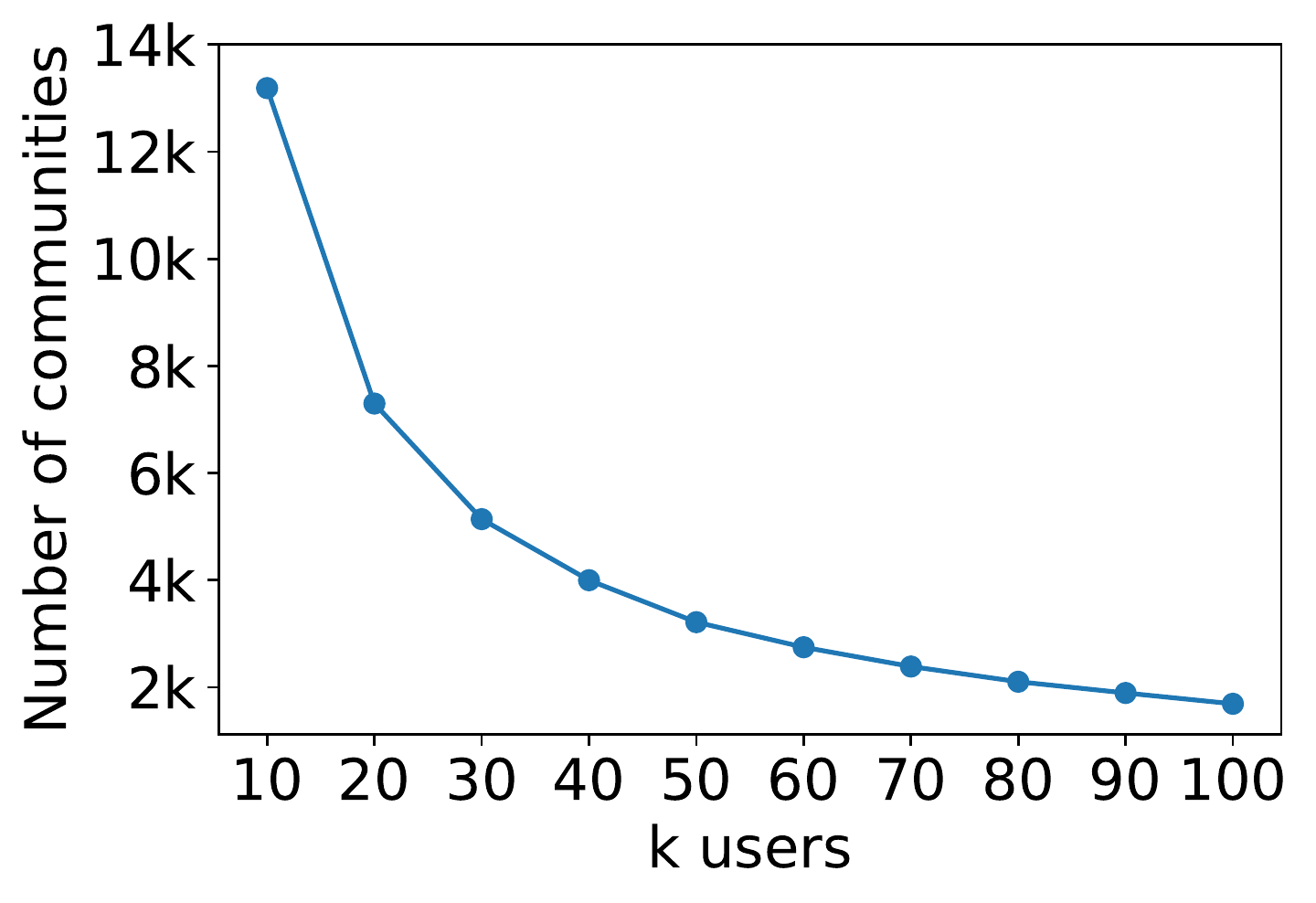}
   \captionof{figure}{Number of communities that attract $k$ users in at most 3 months, with $k$ varying from 10 to 100.}
   \label{fig:num_groups}
\end{figure}

We obtained all posts and comments from subreddits created in 2014 \cite{Baumgartner:2015, Baumgartner:2018}. Since our goal is to investigate the relationship between early community behavior and eventual success, we focus on communities that attracted at least $k$ users within 3 months from the time when they were created so that we have sufficient data to measure the community's early behavior. We vary the value of $k$ from 10 to 100 users. 
It follows that communities with less than 10 users within this time frame are not used in our analysis.
Most subreddits exhibit very low activity after creation. Of all the communities created in 2014, only 5\% attract 10 users within 3 months of their creation. Figure~\ref{fig:num_groups} shows the total number of communities created in 2014 that attracted $k$ users in the first 3 months. 
As expected, the number of communities drops with $k$. However, even for relatively large values, such as $k=100$, we will have more than 1,000 communities that pass the threshold. 



%% file: framework.tex
\section{Characterizing Community Success}

Although community success is an essential problem in understanding group dynamics \citep{kraut:2012,Kim:2000:CBW:518514},
success is an overloaded concept and has been defined in many different ways.
For instance, a battery of studies have focused on growth \citep{tan:18,Kairam:2012}, which implies that communities that attract a lot of users are successful.
These studies tend to focus on factors that attract new users such as social influence and 
tightly connect with research on diffusion \citep{Romero:2011:DMI:1963405.1963503,Cheng+etal:2014,rogers:2010,centola:2010,Bakshy:2012:RSN:2187836.2187907,Romero+Tan+Ugander:13}.
Another heavily studied metrics is retention, also known as churn prediction, which examines the likelihood that a user leaves the community after initial participation \citep{Danescu-Niculescu-Mizil:2013:NCO:2488388.2488416,tan2015all,jurgens2017analysis,Dasgupta:2008}. 
A closely related metric to retention is survival, which examines how long meaningful activity in a community is sustained \citep{Kairam:2012,Zhu:2014:IMO:2556288.2557213}.
In practice, Web companies also use daily active users as a metric of success \citep{Kloumann:ProceedingsOfWww:2015,protalinski2014facebook}.

An important open question is how these success metrics relate to each other and whether they capture the same underlying notion of ``success'', simply known as different names, or they capture inherently different types of success.
In this section, we formally define multiple metrics of success and provide a large-scale characterization of these success metrics and their relations on Reddit communities.
 We believe that all these definitions can be appropriate yet {\em different} because success is not uni-dimensional and can differ due to the varying nature of communities. A successful online discussion group on political news likely present different desirable characteristics when compared to a group formed by editors of Wikipedia articles. For a political news group, an important goal is to sustain daily activity in order for the community to stay updated with daily news, while a group of Wikipedia editors may focus on producing high-quality articles, which requires retaining trained users for a long period of time rather than simply growing the number of users.


\subsection{Defining Community Success}
\label{sec:success-measures}

We consider four classes of success measures that should be generally desirable by most communities: (i) growth in the number of members; (ii) retention of members; (iii) long term survival of the community; and (iv) volume of activities within the community. 
In addition, submitting posts and commenting on posts or other comments are two different ways of engaging with a community and they may be valued differently across subreddits \footnote{In fact, some prior studies only focus on posting behavior for this reason \citep{tan:18,tan2015all}.}.
For example, \communityname{KotakuInAction} is dedicated to discussing controversy centered on issues of sexism and progressivism in video game culture and a single post can generate long discussions through thousands of comments. Other sudreddits are more focused on producing content through posts such as subreddits dedicated to breaking news. We thus further distinguish growth and activity in posts and comments.

To define the success measures, we consider a fixed period after $k$ users joined the new community and use month as the unit for measuring community characteristics. 
We define $T_{i}^{\text{posts}}$ and $T_{i}^{\text{com}}$ as the number of posts and comments in the community in month $i$, respectively. We let
$U_i^{all}$, $U_{i}^{\text{posts}}$, $U_{i}^{\text{com}}$ be all active users, users who posted, and users who commented in the community in month $i$, respectively. Finally, we let
$U^k$ be the initial $k$ users who participated in the community and $U_{i \rightarrow i+1}$ be the subset of users that posted or commented in the community in month $i$ and posted or commented again in the community in month ${i+1}$.

We then formalize our success measures in the following:

\begin{itemize}
\item \textbf{Growth} refers to the number of new users that joined the community in the one year window after the community reached $k$ users.
We consider two types of growth, growth in number of new users that commented (commenters) in the following year, $G^{\text{com}} = |\bigcup_{i=1}^{12} U_{i}^{\text{com}}|$, and growth in the number of users that posted 
in the following year, $G^{\text{posters}} = |\bigcup_{i=1}^{12} U_{i}^{\text{posts}}|$.

\item \textbf{User retention} refers to the average monthly retention rate in the first year after reaching $k$, $R = \frac{1}{12} \sum_{i=1}^{12} \frac{|U_{i \rightarrow i+1}|}{|U_i^{all}|}$.


\item \textbf{Long term survival of a community} is measured by computing the fraction of activity in the final part of our dataset for each community. That is, instead of looking at the following year, we investigate if the community ``died'', or stopped being active after two years. We measure the percentage of activities in the last 3 months of 24 months time window after the community size reaches $k$,
$\frac{\sum_{i=22}^{24} (T_{i}^{\text{posts}} + T_{i}^{\text{com}})}{\sum_{i=1}^{24} (T_{i}^{\text{posts}} + T_{i}^{\text{com}})}$. Our intuition is that if the volume of activity during these final months is very small relative to the overall activity, then the community is not surviving in the long term. 

\item \textbf{Volume of activities} is divided in two types --- the average number of posts in the first year after reaching $k$ users,  $\frac{1}{12}\sum_{i=1}^{12}T_{i}^{\text{posts}}$, and the average number of comments in the first year after reaching $k$ users,  $\frac{1}{12}\sum_{i=1}^{12}T_{i}^{\text{com}}$. 
\end{itemize}


\subsection{Correlation between Success Metrics}
\label{sub:successMeasures}

\begin{figure*}[th]
\centering
\begin{subfigure}[t]{0.55\textwidth}
\centering
\includegraphics[width=\linewidth]{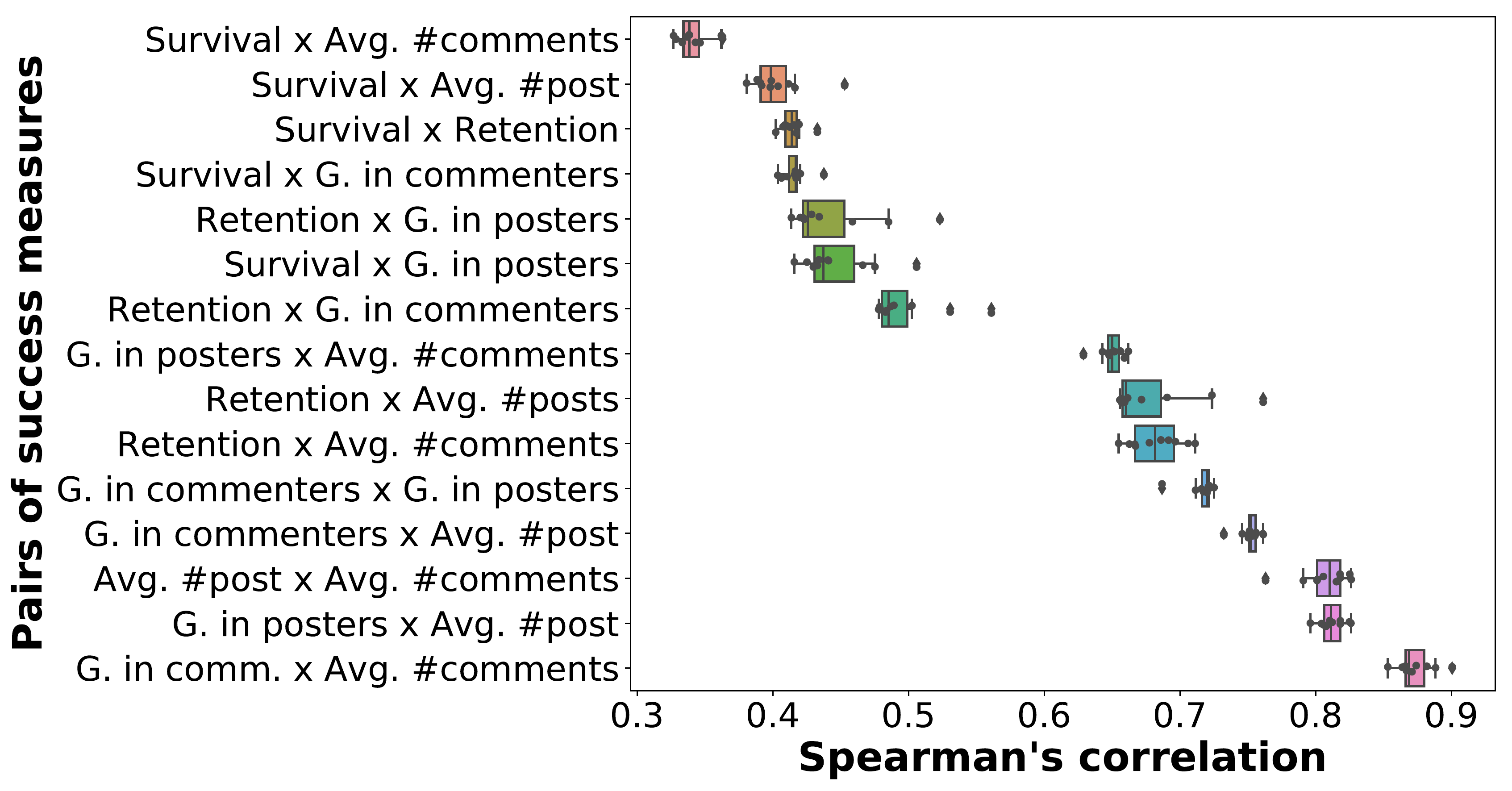}
\caption{. Each box plot represents the distribution of average Spearman's correlation coefficient between two success measures (the number of samples in a box is 10, corresponding to $k=10, 20, \ldots, 100$). Although all the pairwise correlations between success measures are positive, there is great variance with regard to how correlated they are, ranging from 0.34 (between survival and average \#comments) to 0.86 (between growth in commenters and average \#comments).} \label{fig:correlations-A}
\end{subfigure}\hspace*{\fill}
\qquad
\begin{subfigure}[t]{0.43\textwidth}
\centering
\includegraphics[width=\linewidth]{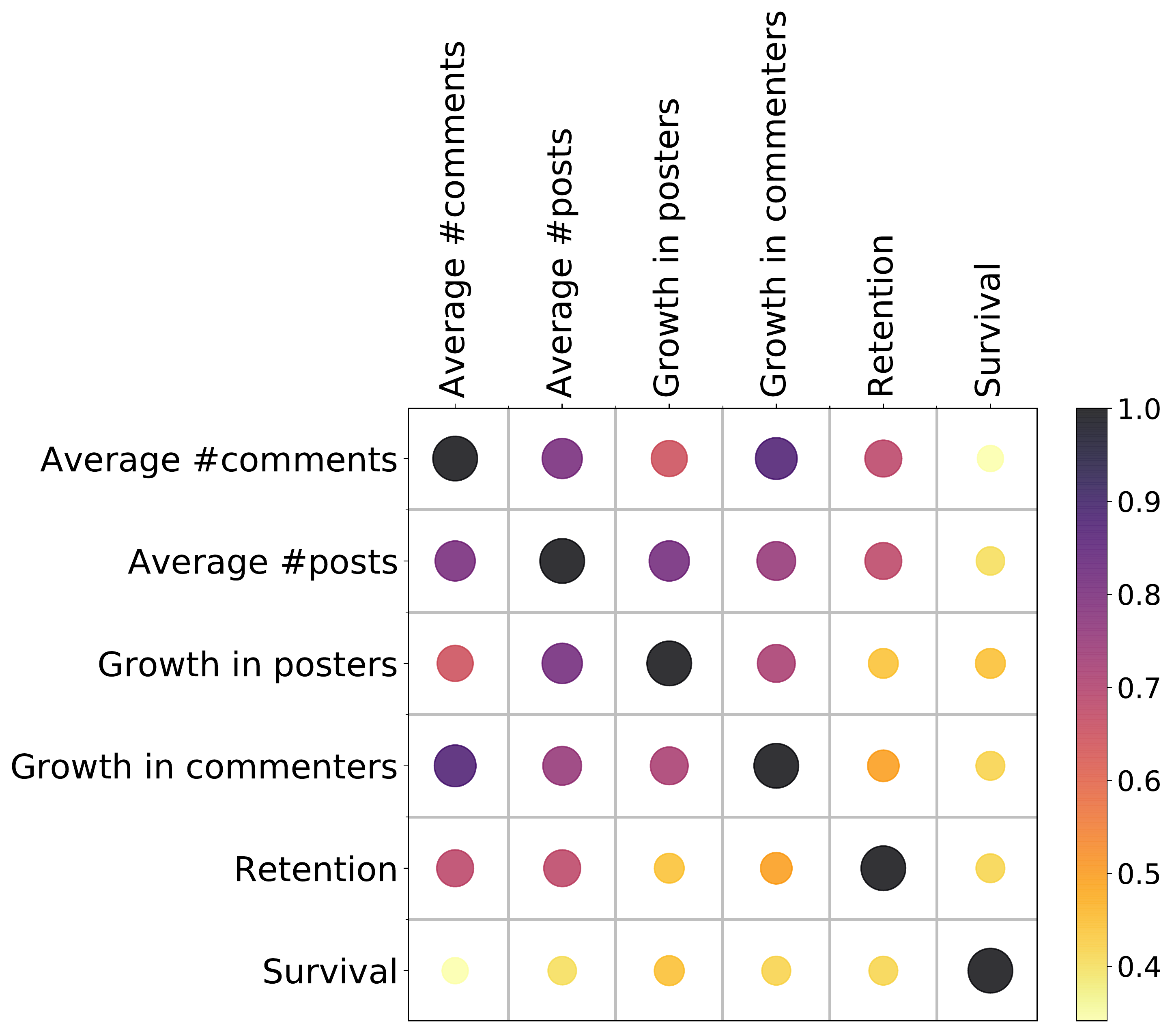}
\caption{Pairwise Spearman correlation between the success measures. Measures related to volume of activities and growth present highest correlations. While survival and retention present the lowest correlations.} \label{fig:correlations-B}
\end{subfigure}
\caption{Spearman's correlation between success measures. We chose Spearman's correlation because it is non-parametric, we also experimented with Kendall's Tau correlation and the results were similar.}
\label{fig:correlations}
\end{figure*}


The premise for this paper is that online communities can be successful in different ways. Thus, our first task is to test this premise by exploring the relationship between the different success measures we identified. A high correlation between all of our success measures would indicate that, in reality, there is a single type of success that achieves all the desirable qualities we described. 

We begin by computing all pairwise Spearman's correlations among the set of success measures,\footnote{We obtain very similar results when we apply Kendall rank correlation \citep{kendall1938new}.} which are shown in figure \ref{fig:correlations-A} in ascending order. Since we can compute the correlation between success measures for each value of $k$, the figure shows a box plot for each pair of success measures for the distribution across $k$. We observe that, while all pairs of success measures have a positive correlation, many pairs have a relatively small correlations --- especially considering that all our success measures depend on users posting content on these communities.

We also present in Figure \ref{fig:correlations-B} a heatmap with the average Spearman's correlation between the success measures. We find that survival has among the lowest correlation when paired with another success measure --- with all such pairs exhibiting a correlation of at most 0.5. This suggests that growth, retention, and high volume of activities do not necessarily imply survival. Similarly, retention exhibits correlations of less than 0.7 with other success measures, except for average \#comments that presented a correlation of 0.75. Correlations are highest among measures of growth and volume of activity, which is not surprising considering that a community with many members will necessarily have a high number of posts and comments since membership is established by activity. The opposite does not necessarily hold. A community could have a small number of very active members. 

These results suggest that indeed, there is significant diversity in the ways that communities achieve success. While some communities can grow and attract a large number of members, others can remain small but succeed by having a set of committed users who always return to the community or by exhibiting long term survival. They also suggest that measuring success with a single metric can overlook important ways in which communities achieve success.

\begin{figure}[t]
    \centering
   \includegraphics[width=0.48\textwidth]{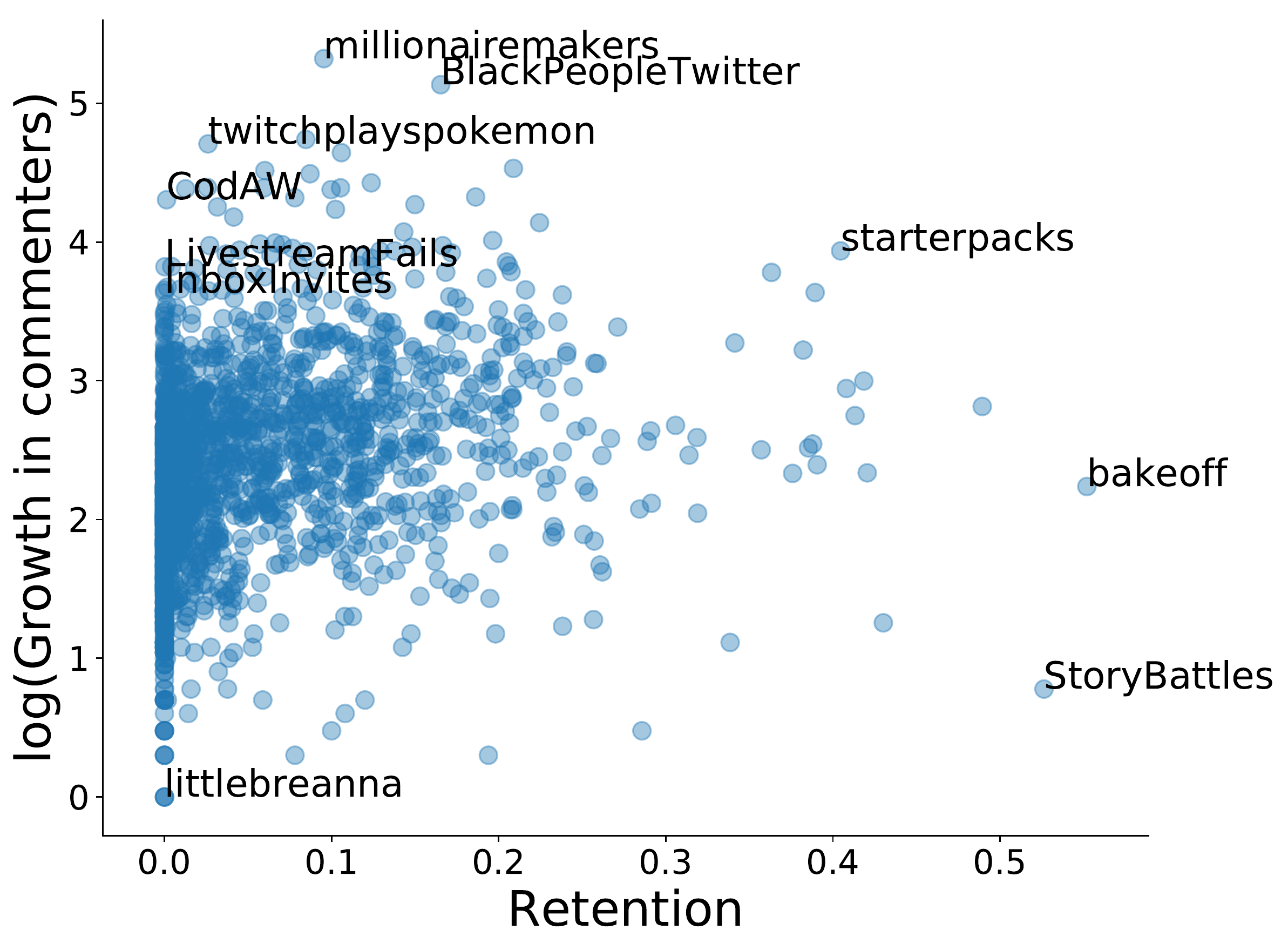}
   \caption{A case study to show the scatter plot between growth in commenters and retention for $k=100$.
   }
   \label{fig:scatter}
\end{figure}

\para{A case study between growth and retention (\figref{fig:scatter}).}
To further understand the differences between success measures, we show the scatter plot of user retention and growth in commenters for all communities that reach $k=100$.
In this scatter plot, successful communities can assume various combinations of values for both dimensions. 
Visually, the growth in commenters can take a wide range of values for communities with zero user retention.
A closer examination of extreme points in this scatter plot reveals interesting examples.
Consider the community \textit{/r/millionairemakers}. In this community, users organize themselves in a ``lottery'', where a user is randomly selected each month and everyone is encouraged to send this user \$1. The winner typically receives between \$1,000 and \$5,000. This community exhibits a large number of commenters because users enter the lottery by leaving a comment. The low user retention might be explained by the fact that winning the lottery is highly unlikely, which may discourage users from participating multiple times. On the other hand, the community \communityname{StoryBattles} --- a subreddit where users can post a picture and the rest of community is asked to write a story about it in the comment section, has very high retention but low growth in commenters. This behavior may be explained by the niche nature of the community, which targets at people interested in story writing. Most Reddit users are probably not interested in story writing, but those who are, return and contribute to this community repeatedly over time. 




%% file: prediction-features.tex
\section{Predictive Features of Success}
\label{sec:features}

Given the relatively low correlation between success measures,
it remains an open question what characteristics of communities predict future community success.
Inspired by existing research and theory in online communities, group dynamics, and organization theory, we defined a set of features to be used later in our success prediction tasks (Section~\ref{sec:experiments}). We focus on features that generalize across different types of communities. We divide these features into six categories which capture different dynamics of the communities. All of our predictive features are based on the actions taken by users between the time the community was created and the time its $k^{\mathrm{th}}$ user arrived. 

\para{Volume and speed of activities.} The first set of features capture both the speed of and volume of activities early during the early stages of the formation of the community, which have been shown to predict future success \cite{kutcher:2014}. It includes, the number of posters, the number of commenters, the date the community was created,
the number of posts, the median number of replies received per posts, the median number of comments and posts per user, the number of days that the community took to attract $k$ members. It also includes the speed at which comments and posts arrive  --- average number of days between posts and comments), which constitute strong baselines used in previous studies \citep{Kairam:2012, cheng:2014-a}.

\para{Distribution of activities.} Another important set of features we explore is related to how the content produced in the community is spread among users and over time (i.e. is a small number of users responsible for producing a large fraction of the activities in the community? Is the majority of the content concentrated in a certain point in time in a bursty fashion?). We capture this spread through the Gini coefficient~\cite{Dorfman:1979}, which is a measure of statistical dispersion commonly used to measure inequality. A Gini coefficient (Equation~\ref{gini_eqn}) of 0 indicates perfect equality where all members produced the same amount of content, while 1 indicates perfect inequality where a single individual generated all posts and comments. We compute the Gini coefficient of the number of posts and comments per users and the Gini coefficient of days between posts and comments. A high Gini coefficient in the number of posts or comments per user is indicative of a small number of community members who are committed to the community enough to produce a large fraction of its content. Such a small set of committed users could ensure the long term success of the community through committed leadership or suggest its demise due to the lack of participation by other community members. 

\begin{equation} \label{gini_eqn}
Gini = \frac{\sum_{i=1}^{n}\sum_{j=1}^{n}|x_i - x_j|}{2n\sum_{i=1}^{n}|x_i|}
\end{equation}

\para{User composition.} The third set of features accounts for the users' activities \emph{prior to joining the community}. It includes the median and standard deviation of users' score ($upvotes - downvotes$) received in posts and comments, the median and standard deviation of the number of users' prior activities and the median number of days users have been active on Reddit. Except for the number of days on the site, we limit this set of features to one month prior to joining the new community. We also measure the fraction of new users who do not have any previous community membership, which includes both new users on Reddit and Reddit users who did not post or comment in the one month before posting or commenting in the new community. As older, more experienced, and more successful users (with respect of upvotes and downvotes) tend to have more experience participating in communities, our hypothesis is that communities that attract more experienced users can benefit as those users were exposed to other communities norms, thus they might produce higher quality content and consequently help the community to achieve its goals. Previous studies have shown that the experience and expertise levels of founders is a good predictor of the success of online communities and offline firms \citep{kraut:2014, davidsson:2003}.


\para{Linguistic style.} The next set of features capture how the language in the content created in the communities can help understand the desirable characteristics of a community. Previous studies  have revealed important stylistic changes in user-written language as users develop a sense of belonging to a community \cite{hamilton:2017, chung:2007, Danescu:2013, tan2015all} and how the sentiment present in the community's feedback affects the likelihood that users will remain engaged \citep{Cunha:2016}. Language can also provide an important signal of users satisfaction, which is a key factor for successful communities, since satisfied users are more likely to contribute to the community and stay engaged \citep{matthews:2015}. First, we compute features that capture the length of the content created in the new community: median post, titles, and comments length and the size of the vocabulary used in the community, i.e. the number of unique words, after pre-processing the content produced by the first \textit{k} users, divided by the number of posts and comments. Next, motivated by studies of socialization and engagement in online communities \cite{Danescu:2013, nguyen:2011, hamilton:2017}, we measure the distribution of categories from the psychological lexicon LIWC \citep{pennebaker:2015, Arguello:2006}~\footnote{http://www.liwc.net}, which measures the various emotional, cognitive, and structural components present in text.
Lastly, inspired by research on corporate success that has found that personality traits are predictive of employees engagement, which boosts the organizational productivity \cite{Shukla:2014, Ziapour:2015, Judge:1999}, we also include the Big Five personality traits \footnote{To compute the Big five personality traits we used the IBM Watson Personality Insights API (https://www.ibm.com/watson/services/personality-insights/).
}: Extraversion, Conscientiousness, Neuroticism, Agreeableness and Openness to Experience \citep{goldberg:1990}. These five domains have been shown to be good at predicting behavioral patterns, such as well-being and mental health, job performance and marital relations \citep{gerlach:2018}. We compute all measures separately for posts and comments.  

 \para{Social networks.} We also include in our analysis a set of features that represent the structure of communication and information exchange among the users in a community. We construct a communication network among early members and extract a variety of features that describe its structure. Network theory has received considerable attention when studying community success. We represent the social interaction among users as an undirected graph $G(V, E)$, where $V$ is the set of vertices and $E$ is the set of edges between a pair of vertices. Each vertex $v \in V$ represent a user in the community, and an undirected edge $e(i, j) \in E$ exists between user $v_i$ and user $v_j$ if user $v_i$ has replied to user $v_j$ or $v_j$ has replied to $v_i$ in a thread. Our choice for undirected edges is based on the assumption that when a post or a comment receives a reply, both users involved are exposed to each other's content. A user \textit{i} must first make a post and a user \textit{j} must read it in order to reply. Once \textit{j} replies, we assume that \textit{i} sees the reply. Thus, we consider that \textit{i} and \textit{j} interacted with each other. 
 
 
We then compute the following network measures: transitivity, average clustering coefficient, density, fraction of users in the largest component and fraction singletons (i.e. fraction of isolated nodes in the graph) \citep{bonacich:1987}. Density, fraction of users in the largest component, and fraction of singletons measure the extent to which the network is well-connected rather than fragmented into many disconnected pieces.  
In a fragmented network, users are not engaging with each other enough to form a large structural community, which could lead to polarization and structural holes \cite{burt2009structural}. 
Transitivity and average clustering coefficient measure the extent to which users have a tendency for sharing connections (or forming triangles). Social networks with high clustering are known to facilitate trust and social capital \cite{coleman1988social}, which is important for community building. Several studies have investigated the relationship between network structure and how the groups change over time \cite{Kairam:2012, tan:18, granovetter:1977, Backstrom:2006, centola:2010, platt:18, romero2016social}. They show that social features are good predictors of communities desirable properties, with special attention given to groups' growth, often growth is treated as a process similar to the diffusion of innovations, where joining a new group is analogous to adopting an innovation. 
 
 Finally, we also measure the fraction of posts and comments that received at least one reply as it was shown that community feedback has a positive effect over users likelihood to return to the community \cite{cheng:2014-b, Cunha:2016, Cunha:2017}.

\para{Parents communities.} The last set of features aims to reveal the emergence of communities through the relationship of the communities that early members were participating before joining the new community. Following Tan's work \cite{tan:18}, we employ features derived from the genealogy graph of parents communities. In the genealogy graph, we consider an edge between two parent communities if they share at least two early members. Then we compute density, transitivity, and number of parents communities. We also include the maximum, minimum and standard deviation of the size of parent communities measured by the log of the number of members one month prior the creation of the new community). Additionally, we include the Gini coefficient of the number of users from the parent communities.
Lastly we measure the median distance between the language model of the focal community and its parents communities (i.e., the median language distance (cross-entropy) between the new community and the parents communities). We consider large language distance as a sign of diversity as \citet{uzzi:2013} has shown that atypical combination is related to scientific impact. 
Note that communities in \citet{tan:18} are only based on posts, while we use both posts and comments.


%% file: experiment.tex
\section{Prediction Results}
\label{sec:experiments}

We now test how well our features described in \secref{sec:features} predict the different measures of success described in \secref{sec:success-measures}. 

\subsection{Experimental Setup}

The prediction for each success measure is framed as a binary task to predict whether that measure for a community will exceed the median value of that measure for all communities considered.  This prediction setup naturally leads to a balanced classification task for all success measures.
It also allows us to ask how the predictability of communities desirable properties varies over the range from small to large number of members.

We train separate logistic regression classifiers with $L2$ regularization 
for each feature category as well as a single model with the combination of all features.  
%
For each classification task, we randomly split the data for training (80\%) and testing  (20\%). The feature values are standardized to make regression coefficients comparable across features from different scales. 
Models are then evaluated using AUC; a random baseline achieves an AUC score of 0.5. 
We grid search the best regularization hyperparameter using 10-fold cross-validation on the training set.  
We do not report statistical significance for regression coefficients due to their opaque interpretation when using the 
biased estimates
\cite{Goeman:2012}.

Models are trained on the
Reddit communities that were created in 2014 and reached $k$ users within three months, where the number of users is measured by the total number of unique users who posted or commented on the community.  To capture differences in behavioral information at different stages of a community's early life, we vary $k$ in 10-user increments from 10 to 100.  Note that fewer communities reach 100 users than 10 users, so the total number of communities that we consider decreases with $k$ (see Figure~\ref{fig:num_groups}).
We compute the median value of each success measure for each $k$ separately to determine the label of each community (positive if its success measure exceeds the median).

\subsection{Can Success Be Predicted?}

Our experiments show that success can indeed be predicted:
the best median AUC score for each task is at least 0.72 when predicting with the combination of all features. In the best case, the median AUC score is 0.84 when predicting users retention.  
The prediction performance results for each model are shown in Figure~\ref{fig:results}. 

\subsubsection{Community Growth}

Although growth in posters and commenters are intuitively related, models varied surprisingly in their accuracy at predicting each version of growth. In particular, user features and parent communities perform substantially worse in predicting the eventual growth in users who post (Figure~\ref{fig:growth_1year_posts}) than growth in users who comment (Figure~\ref{fig:growth_1year_comments}). 
This may be related to the fact that the majority of early members based on our definition are commenters.
For growth in commenters, the model trained on all available features have a median performance of 0.75, yet models trained on a single set of features perform substantially worse. Among the single-feature category, the distribution of activities is the strongest single set of features with a median model performance of 0.63. The performance gap between using all features and using distribution of activities is statistically significant ($p < 0.001$), suggesting no single model captures all the aspects involved in community growth in commenters.
For growth in posters, volume and speed of activities is the strongest single feature set, which echoes previous results that predict growth only considering information about posts \cite{tan:18, cheng:2014-a, Kairam:2012}.  However, models in previous works excluded other community features, which when combined with volume and speed, substantially enhances our ability to predict both types of growth. 

%

\subsubsection{Users Retention}
Results for the task of predicting retention are shown in Figure~\ref{fig:retention}. Our first observation is that this task appears to be easier to predict than future growth. Indeed, the set of all features combined presents the best performance with a median AUC of 0.84, which is larger than any other success measure. When looking for single set of features, there is a clear best model, with distribution of activities features performing stronger than the rest of the models, displaying an AUC of 0.82. The worst performing model is the users features, which suggests that the users' background is not important for predicting if a user will become ``loyal'' to the community and return to it.

\subsubsection{Survival}

Survival is the most difficult type of success measure to predict from our features. The prediction results are shown in Figure~\ref{fig:survival}. In this task, our model with all features achieves an AUC of 0.72. The single best model is content features with an AUC of 0.67, followed by the users composition features with an AUC of 0.63. This observation further confirms the importance of the content generated in the community to keep users committed to the community. It also suggests that early members' previous experience and characteristics are important to community survival, which resonates with \citet{kraut:2014} that found that founders social capital is a good predictor of online communities survival.

\subsubsection{Activity level}

\begin{figure*}[thp]
\centering

\begin{subfigure}{0.32\textwidth}
\includegraphics[width=\textwidth]{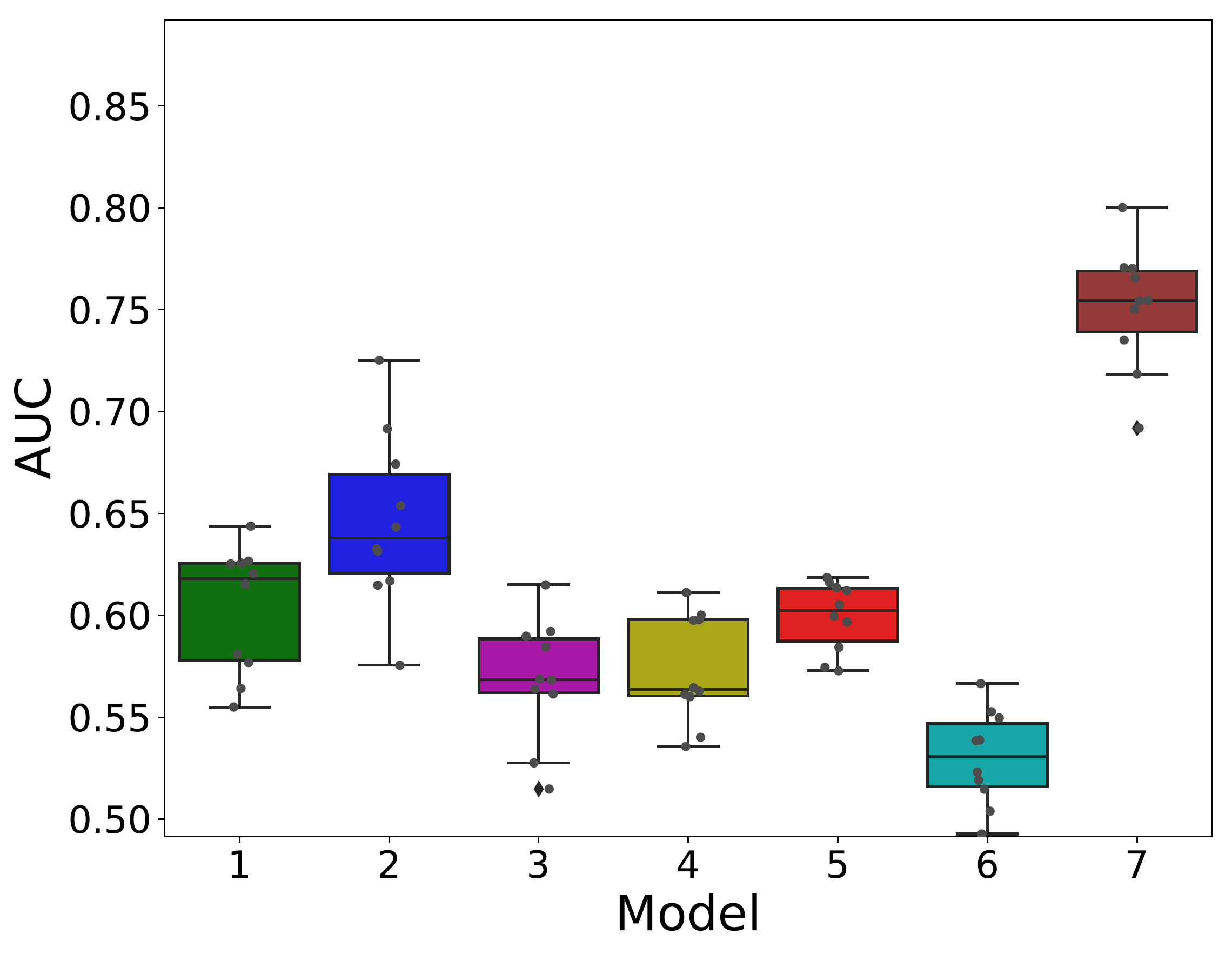}
\caption{Growth in commenters} \label{fig:growth_1year_comments}
\end{subfigure}
\begin{subfigure}{0.32\textwidth}
\includegraphics[width=\textwidth]{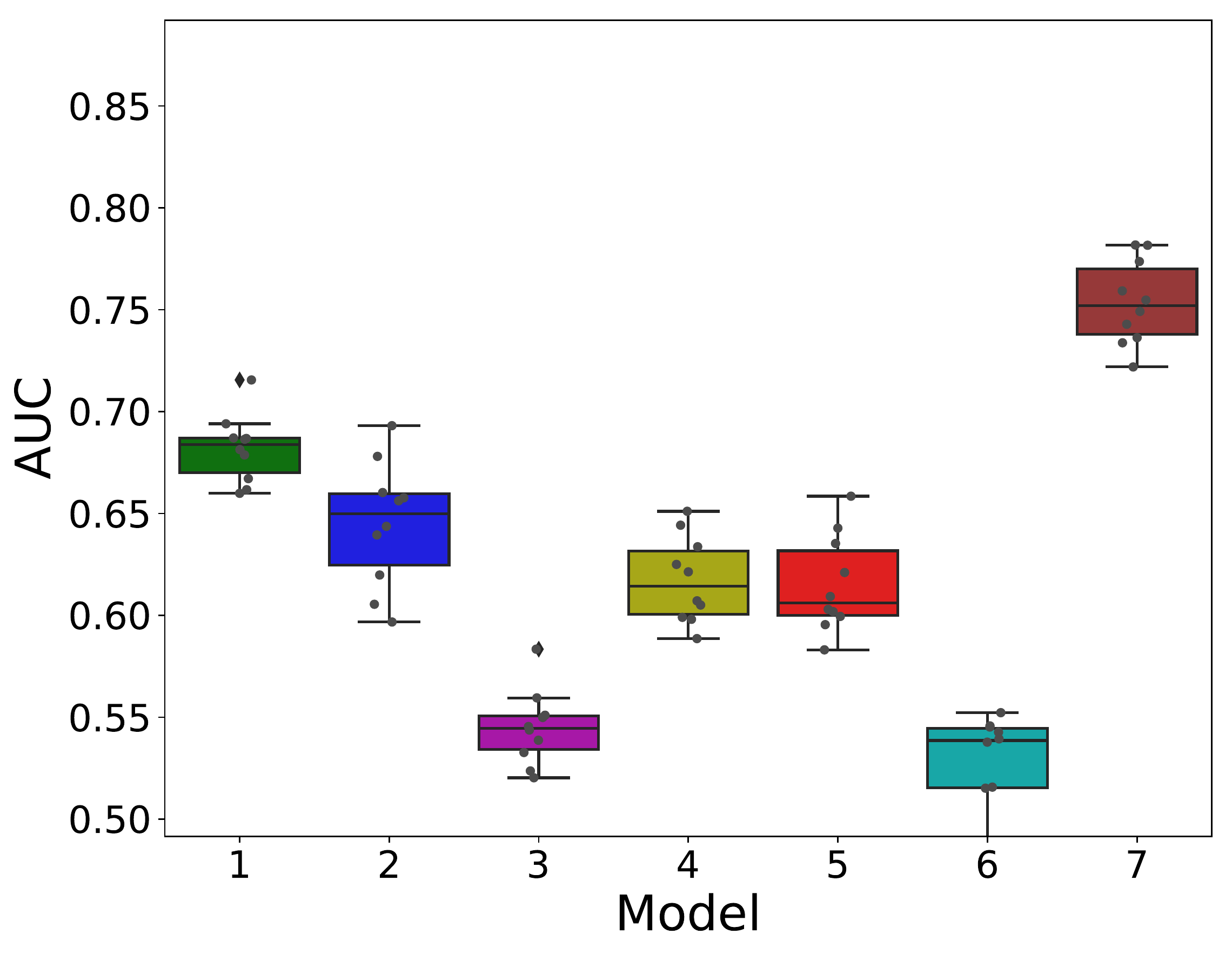}
\caption{Growth in posters} \label{fig:growth_1year_posts}
\end{subfigure}
\begin{subfigure}{0.32\textwidth}
\includegraphics[width=\textwidth]{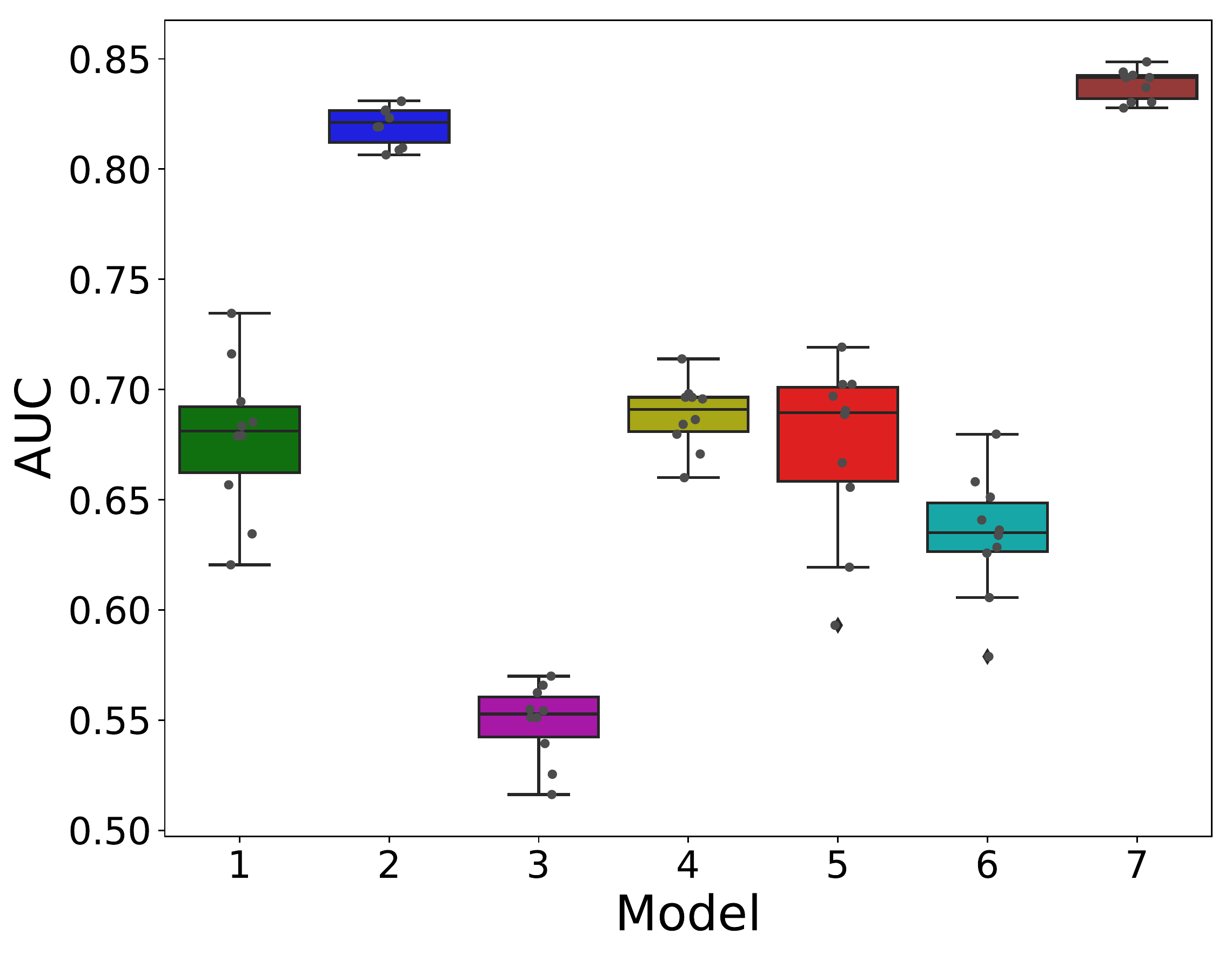}
\caption{Retention} \label{fig:retention}
\end{subfigure}

\begin{subfigure}{0.32\textwidth}
\includegraphics[width=\textwidth]{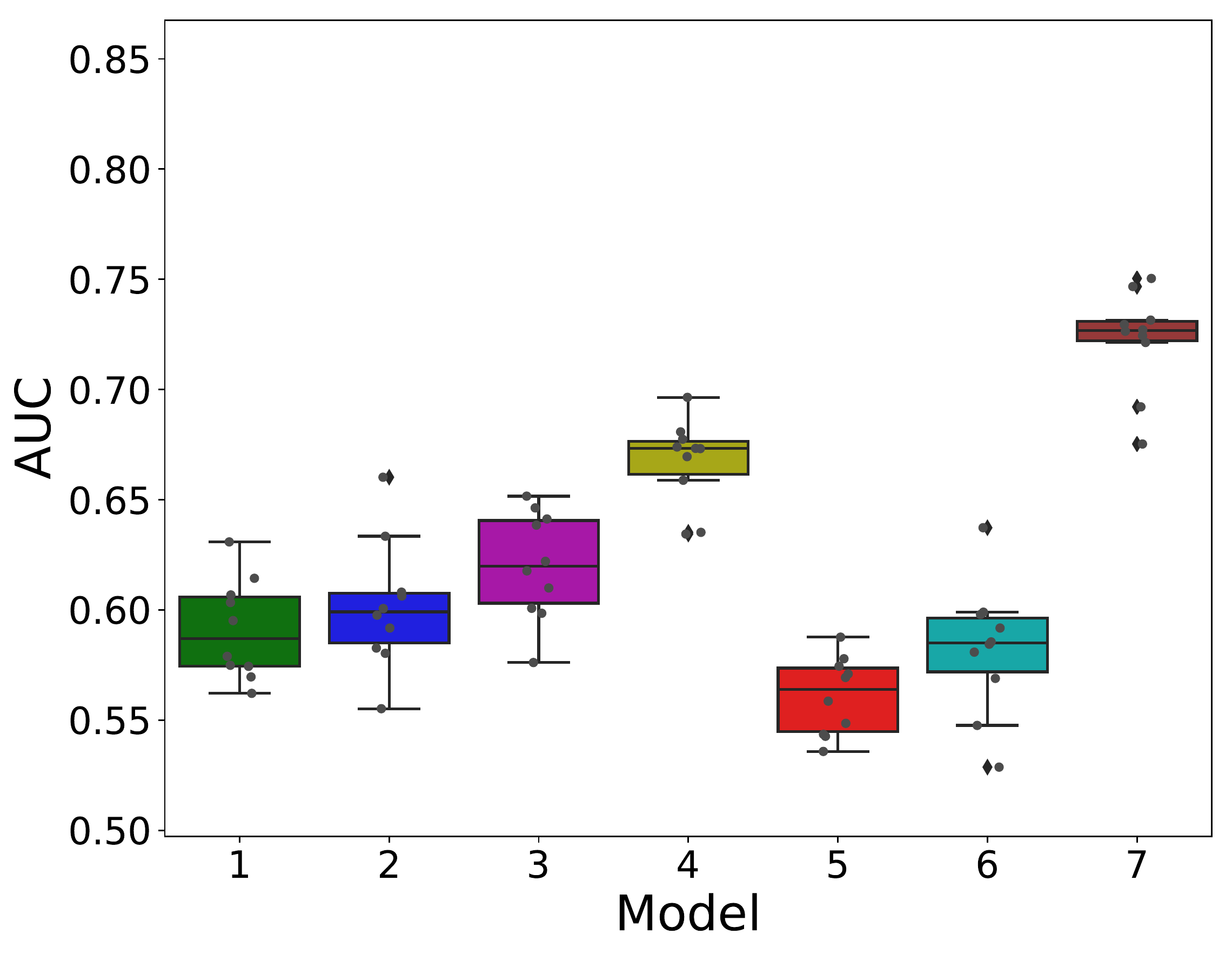}
\caption{Survival} \label{fig:survival}
\end{subfigure}
\begin{subfigure}{0.32\textwidth}
\includegraphics[width=\textwidth]{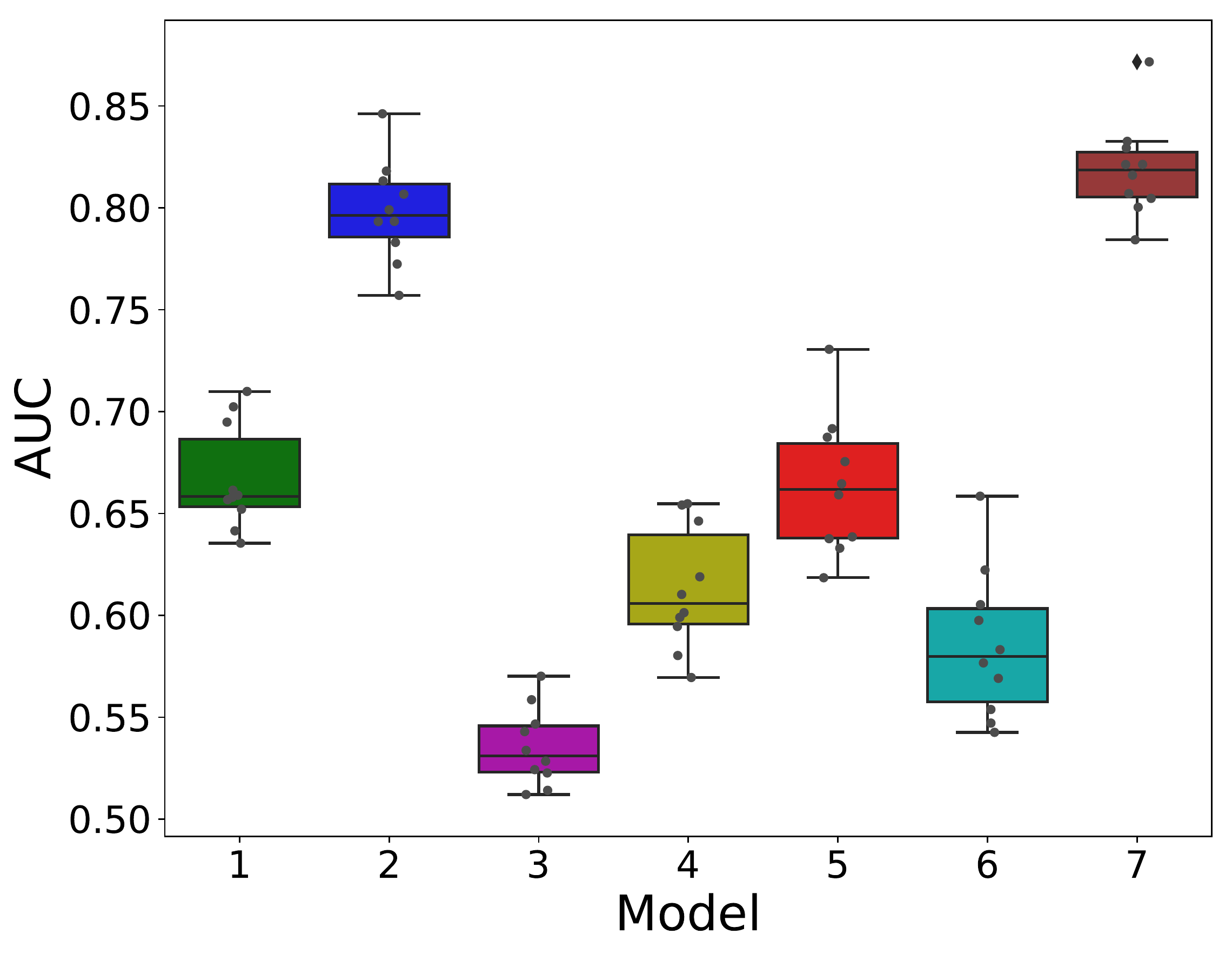}
\caption{Average \#comments} \label{fig:avg_comments}
\end{subfigure}
\begin{subfigure}{0.32\textwidth}
\includegraphics[width=\textwidth]{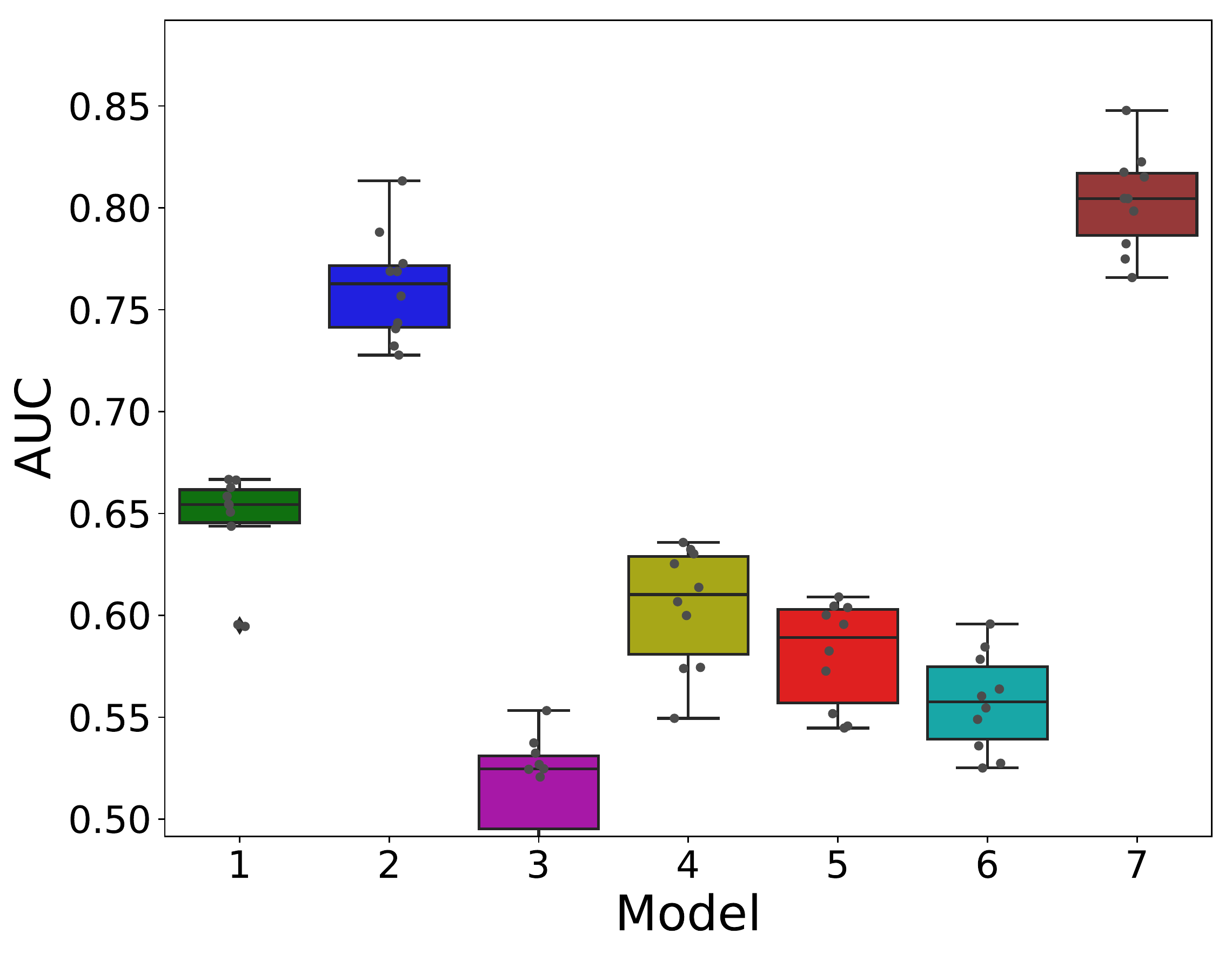}
\caption{Average \#posts} \label{fig:avg_posts}
\end{subfigure}

\begin{subfigure}{0.8\textwidth}
\includegraphics[width=\textwidth]{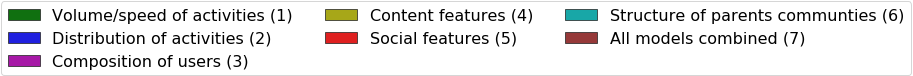}
\end{subfigure}

\centering \caption{Results for the predictive tasks of success measures. Box plots graphically depict the results for $k$ varying from 10 to 100 users. They show that our features sets are predictive of communities' success measures, the model of all features combined presents at least 0.72 AUC for all task.} \label{fig:results}
\end{figure*}

Next, we evaluate the predictive power of our feature sets to predict the future activity level. The results for average \#comments are shown in Figure~\ref{fig:avg_comments} and results for average \#posts are shown in Figure~\ref{fig:avg_posts}. The results for the two tasks are similar, with the model with all features performing slightly better for average \#comments. This can be explained by the performance of the social features model, which presented a better result (0.67 vs 0.58, $p < 0.001$) for the task of predicting average \#comments. The model with all features outperformed the single features models for both tasks, but this difference was only statistically significant for the task of predicting average \#comments (0.80 vs 0.76, $p < 0.001$.). The worst results for single features models were presented by the model with user features, which suggests that the users' history before joining a new community does not play an important role in the frequency of users to post/comment in the future.



\subsection{How Early is Success Predictable?}
    
Each of the box plots in Figure \ref{fig:results} contains the resulting AUC for each value of $k = 10,...,100$. One may expect the performance of the model to increase with $k$, as a higher $k$ represents data from more users to predict the eventual success of the community. Surprisingly, most of the box plots display a small range of variation --- indicating that the performance of the model is in fact not very sensitive to the number of early users we base our predictions on. In particular, the AUC for $k=10$ is not very different from that of $k=100$. Table~\ref{tab:dispersion} shows the standard deviation of the AUCs for all features combined when applied to all success measures. The small deviation values confirm our intuition that indeed the features are predictive since the community's infancy. This result is important for communities maintainers and moderators as they can diagnose their communities in their initial stage and identify ``at risk'' communities.   



\begin{table}[t]
\begin{tabular}{lr}
\toprule
\textbf{Success measure} & \textbf{Std. AUC} \\ \midrule
 Growth in commenters & 0.030 \\ 
 Growth in posters & 0.020 \\ 
 Retention &  0.006 \\ 
 Survival &  0.022 \\ 
 Average \#comments  & 0.023 \\ 
 Average \#posts &  0.024 \\ \bottomrule
\end{tabular}
\caption{Standard deviation of results for the model with all features combined for all success measures.
}
\label{tab:dispersion}
\end{table}
\iffalse \daj{Does this need to be a table?  I'm not sure what is really learned from this at the moment and feel like it could just be said in text.}}\fi



\subsection{Which Features Predict Success?}
\begin{figure*}[thp]
\centering

\begin{subfigure}[t]{0.48\textwidth}
\includegraphics[width=\linewidth]{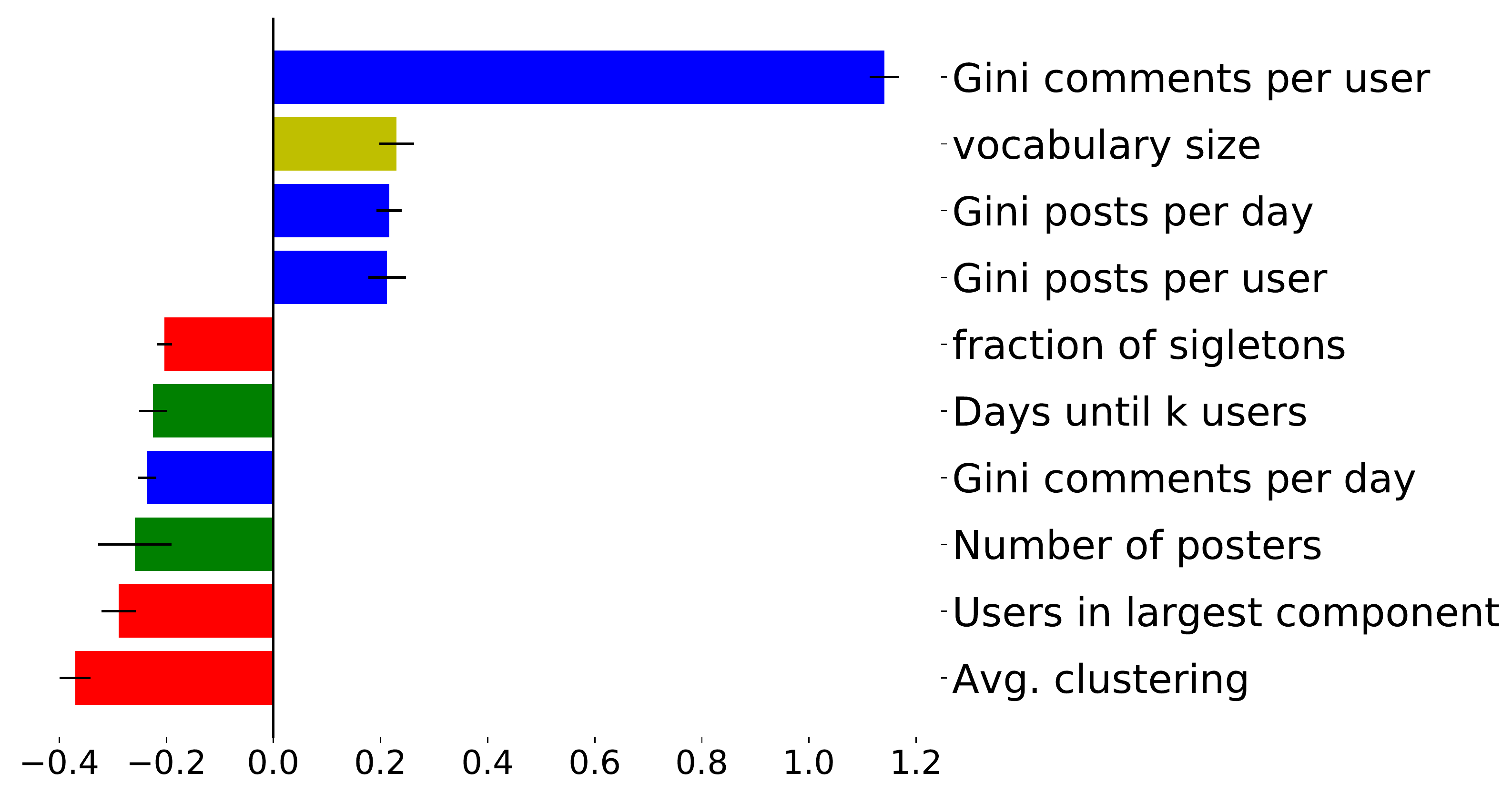}
\caption{Top 10 features for growth in commenters.} \label{fig:top_features_g_comments}
\end{subfigure}
\hspace*{\fill}
\begin{subfigure}[t]{0.48\textwidth}
\includegraphics[width=\linewidth]{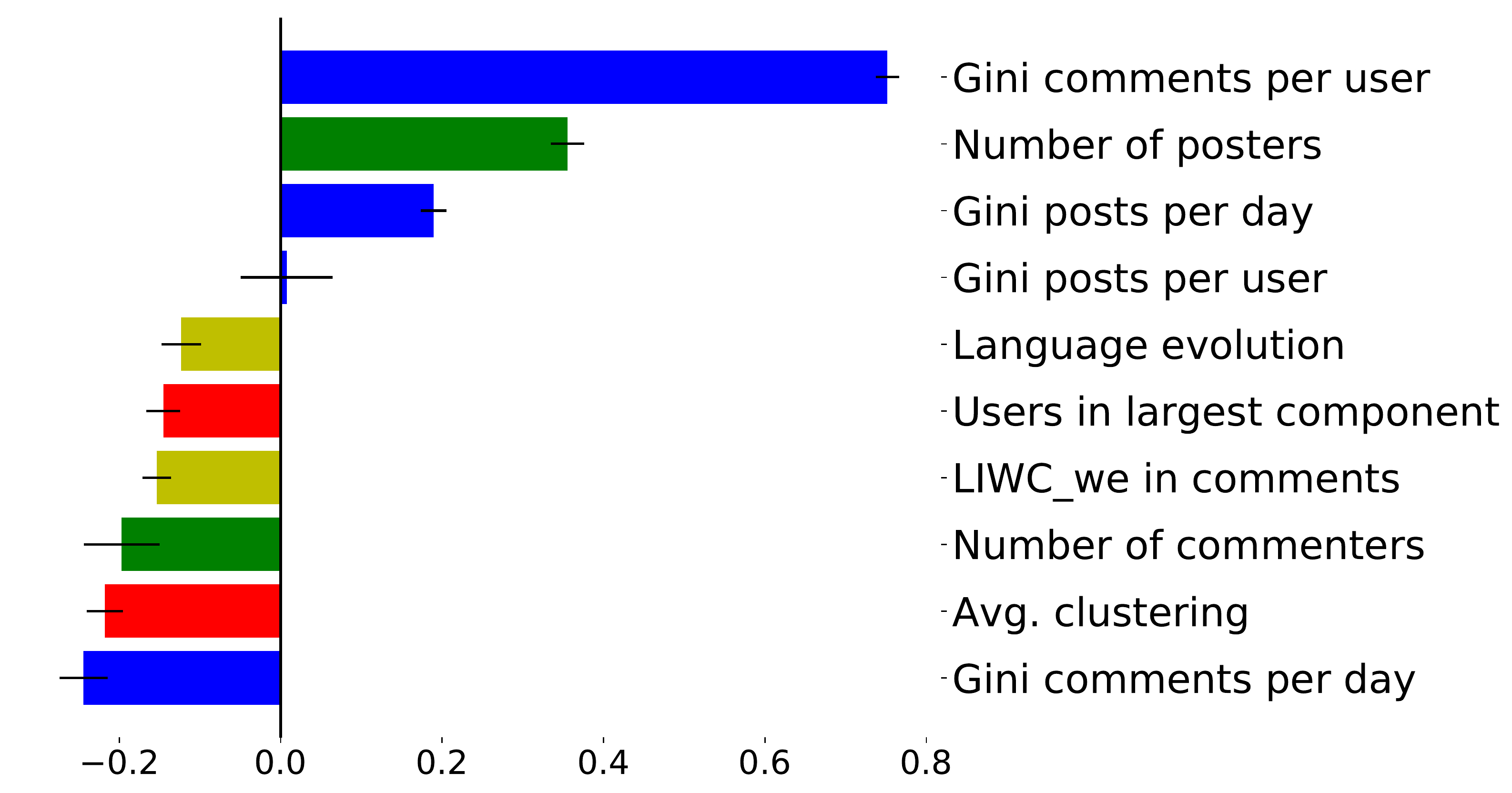}
\caption{Top 10 features for growth in posters.} \label{fig:top_features_g_posts}
\end{subfigure}
\medskip

\begin{subfigure}[t]{0.48\textwidth}
\includegraphics[width=\linewidth]{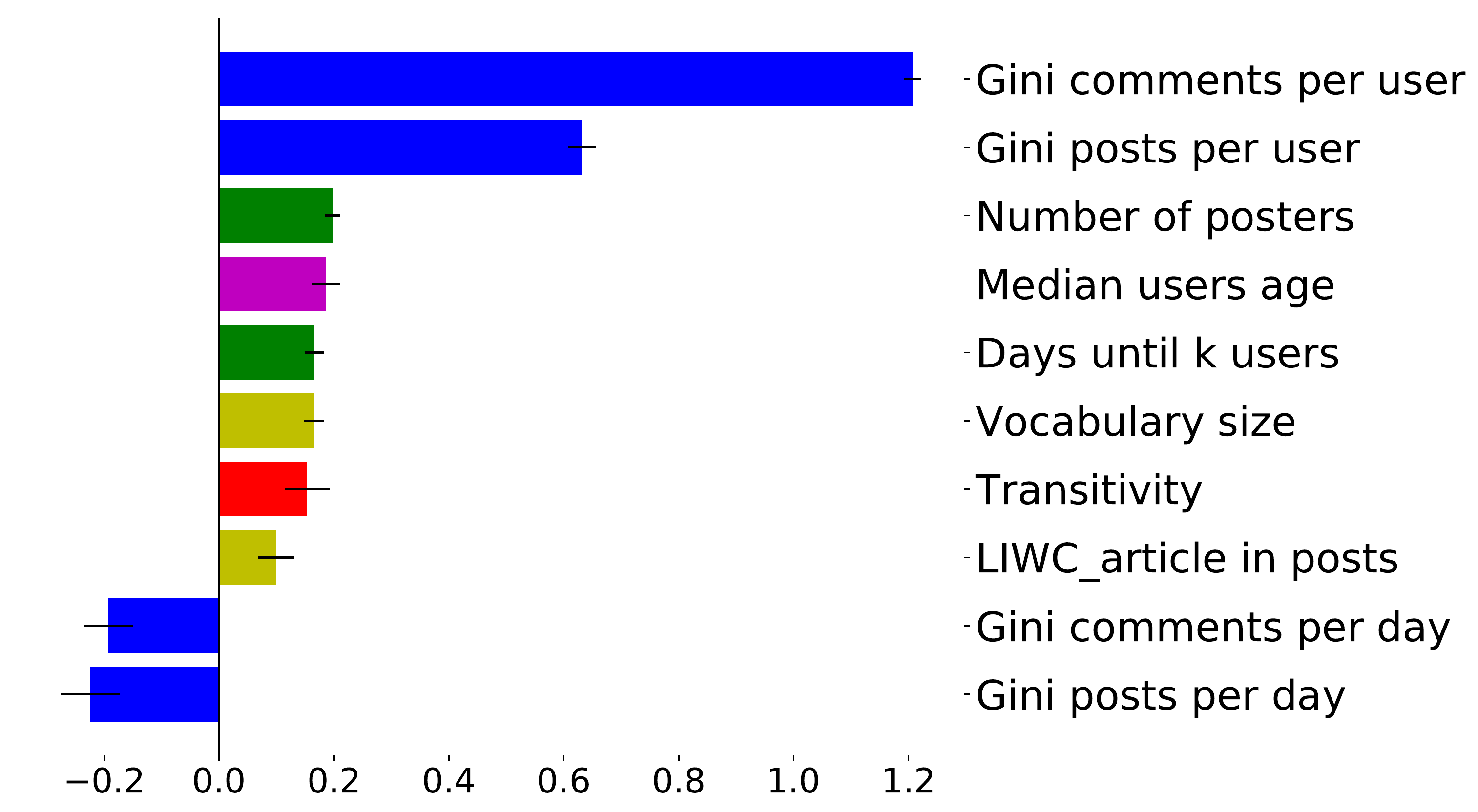}
\caption{Top 10 features for users retention.} \label{fig:top_features_retention}
\end{subfigure}\hspace*{\fill}
\begin{subfigure}[t]{0.48\textwidth}
\includegraphics[width=\linewidth]{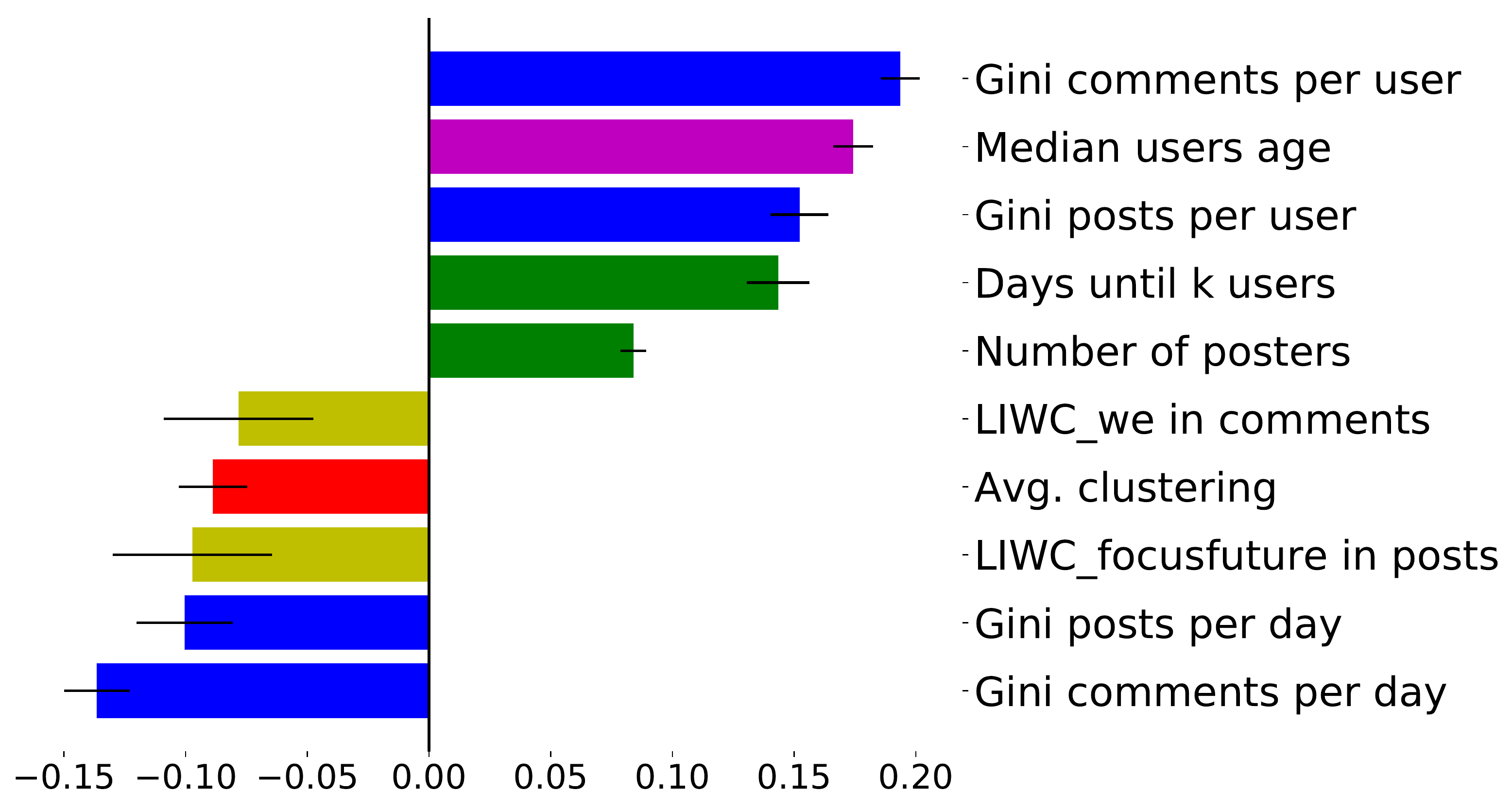}
\caption{Top features for community survival.} \label{fig:top_features_survival}
\end{subfigure}
\medskip
\begin{subfigure}[t]{0.48\textwidth}
\centering\includegraphics[width=\linewidth]{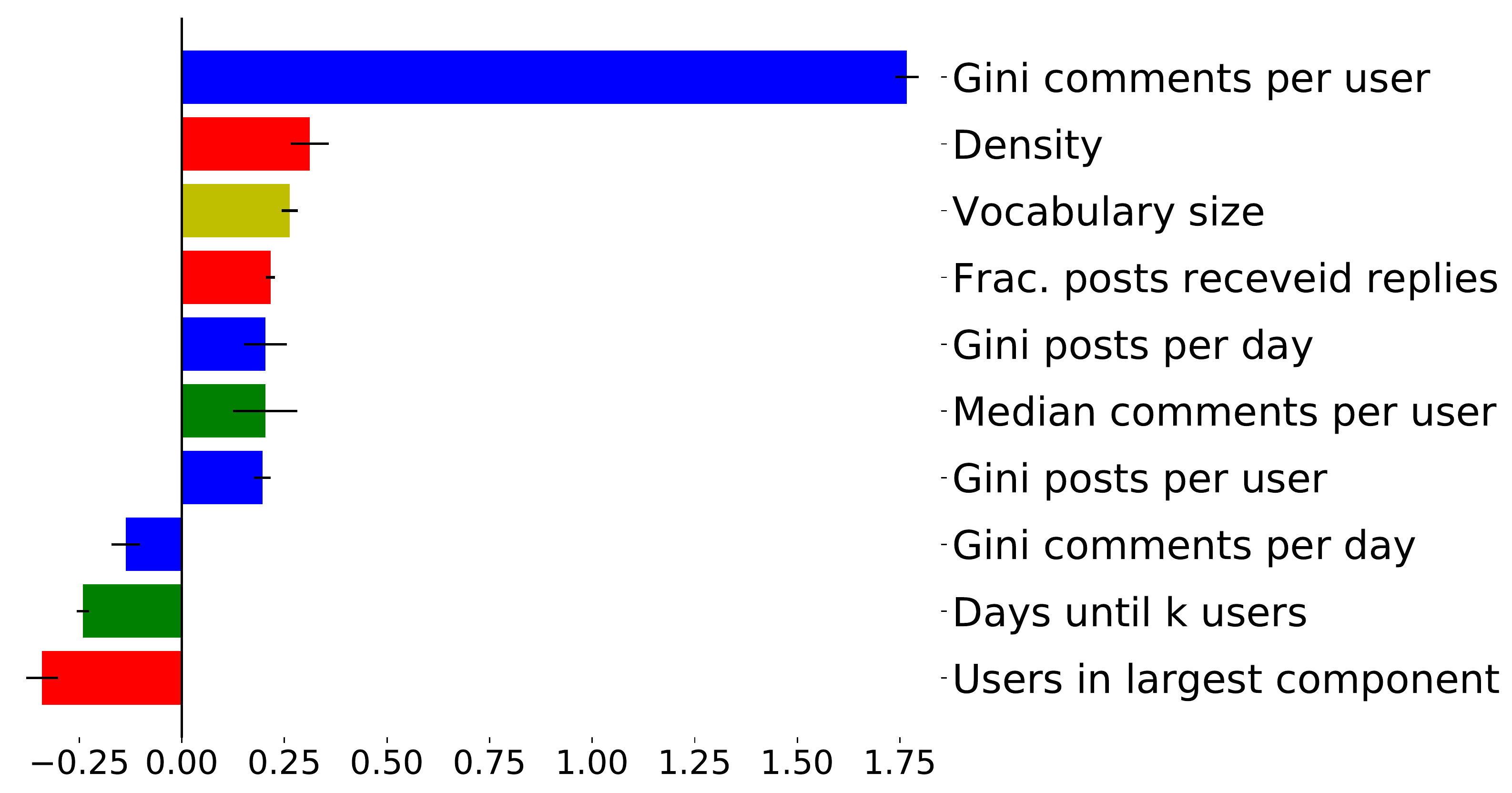}
\caption{Top 10 features for average \#comments.} \label{fig:top_features_avg_comments}
\end{subfigure}\hspace*{\fill}
\begin{subfigure}[t]{0.48\textwidth}
\centering\includegraphics[width=\linewidth]{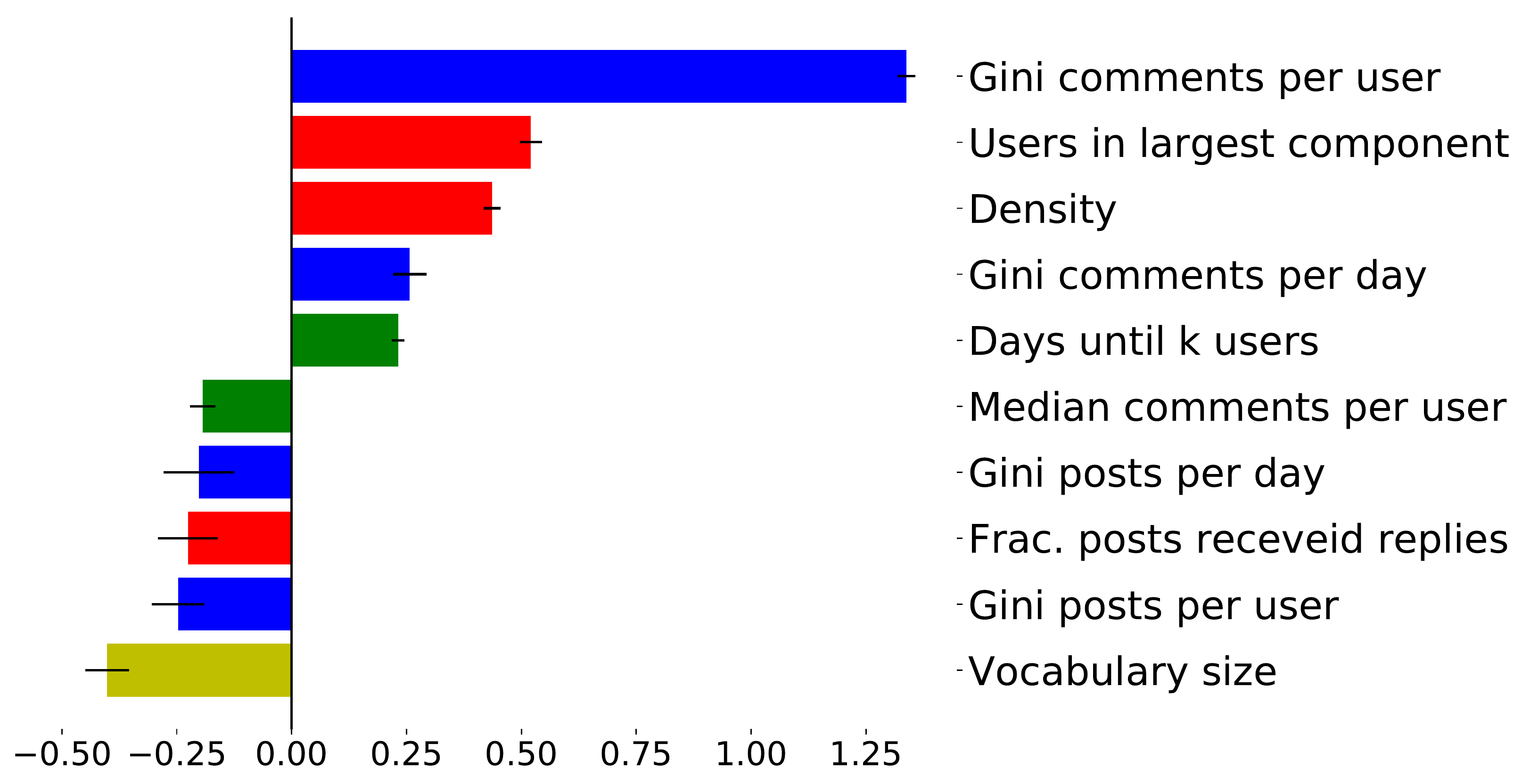}
\caption{Top 10 features for average \#posts.}\label{fig:top_features_avg_posts}
\end{subfigure}

\caption{Top 10 most predictive features for the model with all features combined}
\label{fig:top_features}

\end{figure*}

In this subsection we examine the most predictive features of success when we combine all our features into a single model. Recall that we have a separate model for each value of $k$. For each $k$, we rank the features of the resulting model by the magnitude of its coefficient. We then use the mean reciprocal rank \citep{craswell:2009} of the features over the individual ranks for $k=10,...100$ to generate a single ranking of the features. Figure \ref{fig:top_features} shows the top 10 features and their mean coefficients. 

The most important set of features is the distribution of activities. Its features appeared among the most predictive for all measures of success. It shows that except for average \#posts, higher Gini coefficient of posts and comments per users is positively associated with all success measures, which means that if the initial activities in the community are concentrated in a small number of users, it increases the overall likelihood that the community will be successful. In early stages, when communities do not have many members or much content, they must rely on 
committed users that are more likely to engage and produce high-quality content.




We also find that the time the community takes to reach $k$ users is predictive of all success measures, except growth in posters. However, the direction of the effect of this feature is not consistent across success measures. While a small number of days to reach $k$ users (i.e. fast initial membership growth) increases the likelihood that a community will grow and survive, it decreases the ability of the communities to retain users. These findings resonate with research on the founding of organizations, which shows that fast initial growth predicts longer organization survival \citep{kraut:2014}. However, our results show that this long term survival can come at the cost of lower retention. Additionally, we note that average \#posts and \#comments are also affected differently by this feature. Though a small number of days to attract $k$ new members predicts an increased average number of comments, but is related to an decreased future average \#posts. Thus, it appears that communities that grow slowly in membership during early stages exhibit higher levels of interactions among users through comments at the cost of lower production of original content through posts. 

Regarding social features, increased average clustering and fraction of users in the largest connected component are negatively associated with growth, while transitivity is positively associated with user retention, which is in accordance with previous results in groups growth \cite{granovetter:1977, Backstrom:2006, Kairam:2012}, where users in very clustered groups tend to persist in the community. However, this property also decreases overall growth and the likelihood to survive longer. In addition, when members are too tightly connected to the largest component, as they are in groups with high transitivity, groups might become too closed, restricting the possibility of gaining new information and attracting new members \cite{Kairam:2012}. Lastly, increased density predicts increased level of activities, which suggests that social interaction predicts future activity level. 

In term of linguistic features, we note an interesting negative association between the use of first-person plural pronouns ``\textit{we}'' and a community's long term survival and growth. 
This finding indicates that users in large communities that survive longer are less likely to assume a collective identity,
which echoes \citet{palla:2007} that suggests a connection between fluid dynamics in membership and a group's long-term survival.
In contrast, findings from sociolinguistics show that loyal members (in terms of user preference and commitment) of communities presented a higher use of first-person plural pronouns ``\textit{we}'' and language that signals collective identity \citep{hamilton:2017, chung:2007,sherblom:1990}.
The diverging effect of ``\textit{we}'' confirms the low correlation between retention and survival, and further demonstrates the multi-faceted nature of success.

%% file: limitations.tex
\section{Discussions}


\subsection{Limitations}

\textbf{Not controlling for effect of a community's topic}. It is possible that communities related to certain topics have different likelihoods to be successful. Although we do not control for the topic of the communities or how its chosen topic fits within the larger ecosystem, we aim to identify a set of features that generalize across topics and types of communities such that our recommendations are general enough and can be applied to any type of new community. 

\textbf{Experiments focus on subreddits created in 2014}. Our entire study is based on 2014 Reddit data, which is late (8 years) into the Reddit ecosystem, when many communities have already been established. While analyzing success of communities in the earlier stage of Reddit could reveal interesting patterns, the analysis would be more challenging as Reddit was changing its platform and evolving rapidly during its early years. Additionally, the large number of potential missing (i.e. deleted users) data from Reddit's earliest years may also introduce biases to the analysis \citep{Gaffney:2018}. However, a replication of our analysis using a much larger time period or studying Reddit alternatives during their formative growth periods \cite{newell2016user} could determine if the predictors of success change over time. 


\textbf{Causality}. In our models, we have interpretable features, but we do not have a causal model to suggest that the presence of a feature \textit{causes} the community to succeed. Applying experimental or other causal inference techniques such as propensity score matching would be a possible line of future research to establish a causal link between community behaviors and  success. 


\textbf{Our analysis only uses posts and comments}. There are other aspects of Reddit communities that may be informative such as stated community norms, moderator behaviors such as banning users or removing posts. However, currently this type of data is either not available or it's difficult to obtain a historical version, which makes their use impractical in our analysis \citep{chandrasekharan:2018}. 



%% file: recommendations.tex
\subsection{Design implications}

New online communities are formed every day. Given that very early behaviors of such communities is strongly predictive of multiple types of success, we 
discuss design implications for new community organizers to increase their likelihood of success. We reiterate the important caveat that our study is purely observational and does not establish causal relations between behaviors and success.

High Gini in posts and comments per user always predict success, regardless of success measure, which results from having a few users that perform most of the activities. This result suggests that having a small group of highly committed groups of participants early on is key to future success. Community organizers need not necessarily worry when not all participants engage frequently, but need to ensure that a sufficient number of members engage at producing content in order to establish a critical mass, which is responsible for helping the community to become self-sustaining and create further growth \cite{granovetter:1978, rogers:2010}. 

Taking longer to attract $k$ members is negatively associated to growth in commenters, survival and average \#comments, but predicts retention and average \#posts. We interpret these results as reflecting two kinds of community goals. For those communities that want to attract many users, stay active longer and have its activities primarily concentrated on discussions, attracting users quickly is important. However, for retaining those users, what matters is getting the right type of users who are likely to remain active, which may take longer initially.  

Transitivity---which is highly correlated to average clustering---is predictive of retention, suggesting that having a close-knit set of participants who bond well with each other is important for retention but not growth, which is consistent with existing literature on community growth \citep{Kairam:2012, Backstrom:2006, granovetter:1977}. Combined with our result for timing, we posit these results may point to the importance of a sense of belonging and identity within a community; users who join a rapidly growing community (low time-to-$k$) may find themselves lost in the mix and slow to accumulate social capital. It also suggests the importance of strong relationships, which take more time to build and are the combination of frequent engagement, deep interaction, and time spent together in the communities. 

The distribution of arrival rates for comments and posts (measured through the Gini coefficient of gaps) point to two diverging recommendations. Bursty commenting followed by long droughts (high Gini in comment gaps) is negatively associated to all measures of success, except average \#posts. Uniform arrival rates for comments leads to more success --- though as the first recommendation suggests, these comments need not come from a large group of contributors. Here, we posit that because comments are a form of interaction and accumulation of social capital, regular arrival rates encourages users to come back frequently and follow up on existing discussions. This also might be explained by the \textit{anticipation and uncertainty of reward mechanism} present in social media, where constant arrival of new content stimulates the production of the hormone Dopamine, a chemical produced in various parts of the brain and controls moods, motivation and sense of reward, which is linked to users' increased presence and activity in social media \citep{soat:2015}. This result suggests that new community organizers should encourage new comments on regular intervals to promote opportunities for others to become involved in the discussion.

%% file: related.tex
\section{Related Work}
\label{sec:related}



A large body of literature investigates how community-level characteristics influence the success of online communities. Here we discuss how the literature defines success and how the early users affect the likelihood of communities becoming successful.

\textbf{Success as the number of members}. The majority of works on community success adopt a very narrow view of how success manifest in communities. Most existing work tries to predict the volume of popularity of communities, such as the growth in the number of members the community. Often, the growth in membership is treated as process similar to the diffusion of innovations, where joining a new group is analogous to adopting an innovation \citep{granovetter:1977, Backstrom:2006, centola:2010, Kairam:2012, tan:18}. Two works present a broader view of success, Kairam \textit{et. al.} investigate how communities social networks predict growth and community survival, their definition of survival is based on the time the community stops to grow rather than stop to produce content, while Ellis {et. al.} \citep{ellis:2016} estimate the health of a community by measuring the retention rate of participants and the period of time the group stays active.

\textbf{On the importance of early members}. Another important line of work focuses on predicting the success of communities is the characteristics of its early members. Inspired by research on offline organizations that demonstrated the importance of the early members for organization success \citep{davidsson:2003}, they investigate whether user characteristics and behavior early in a community's history  predict community success \citep{Zhu:2014:IMO:2556288.2557213,Zhu:2014:SEN:2556288.2557348,tan:18}. Early members are responsible for creating the initial content and norms of the community, acting as recruiters of new members. For example, Kraut and Fiore \citep{kraut:2012} found that human and social capital, such as users' age and experience  with  other communities' early decisions are predictive of success. Their findings suggest that the resources early members bring to the group are important, but they can also be detrimental if the group depends on them too much.

Our work contributes to prior research in two major aspects. First, we shed light over the idea that success is a complex concept and can be described in multiple dimensions that depend on the goal of a community. We thus investigate success across four desirable properties. Second, we present a set of features that represent the communities' behaviors and can generalize across the different types of communities. Finally, we show that success can be predicted and  reveal which features are the most important for each success definition.

%% file: conclusion.tex
\section{Conclusions}
\label{sec:conclusion}

New groups are created frequently in online platforms. What does it take for a new group to succeed and how do we measure success?  We answer these  questions using a large-scale study on tens of thousands of groups on Reddit. Our work offers two main contributions.  First, we quantify success using four measures and show, surprisingly, that success according to one measures is not necessary met with success with the other three. Instead, while the measures are positively correlated, groups succeed in their own way, in part due to the diversity of \textit{why} a  group forms.  A small community may thrive through consistent posting, while never growing large, whereas another group may expand to a massive number of users, with only a few core people participating.    
Second, we show that the future success of a group can be predicted along each of the four measures. Drawing upon prior work and theories of group organization, we quantify six different group behaviors to test their predictiveness of success. Our results show that no single behavior drives a group to be successful in each dimension.  For example, while retention and growth are well-predicted by the inequality of creating content, group longevity is best predicted by the linguistic style of the content. Further, predictive accuracy is enhanced when combining all behaviors, suggesting that each behavior plays a role in the eventual success (or failure) of a community.  

Together, our results point to the complex roles groups serve online: each group is created with a different purpose and the success of that group is dependent in part on cultivating the types of behaviors needed. Our results also provide practical advice for individuals wanting to start their own new community online.